\documentstyle[12pt]{article}

\newcommand{\beq }{\begin{equation}}
\newcommand{\enq}{\end{equation}}
\newcommand{\lan}{\langle}
\newcommand{\ran}{\rangle}
\newcommand{\eqn}[1]{eq.(\ref{#1})}
\newcommand{\strutje}{\rule[-1.5mm]{0mm}{5mm}}

\begin{document}

\begin{titlepage}
\begin{flushright}
\noindent September 1993 \hfill  KUL-TF-93/41\\
$\strutje$ \hfill   hep-th/9309099\\
\end{flushright}

\vfill
\begin{center}
{\large\bf Chiral Quantization on a Group Manifold}\\
\vspace{10mm}
{\bf \centerline{ Zbigniew Hasiewicz
\footnote{Onderzoeker IIKW, Belgium.
Supported in part by KBN Grant 2 00 95 91 01. } ,
Przemys\l{aw} Siemion
\footnote{On leave from IFT, University of Wroc\l{a}w, Poland}, and
Walter Troost\footnote{Bevoegdverklaard Navorser NFWO, Belgium}}}
 \vskip .3cm
{\baselineskip = 12pt
\centerline{\sl{Instituut voor Theoretische Fysica }}
\centerline{\sl{Katholieke Universiteit Leuven }}
\centerline{\sl{Celestijnenlaan 200D     }}
\centerline{\sl{B-3001 Leuven , Belgium}}}
\end{center}
\vfill
\begin{center}
{\bf Abstract}
\end{center}
\begin{quote}
\small
    The phase space of a particle on a group manifold can be split in
left and right sectors, in close analogy with the chiral sectors in Wess
Zumino Witten models. We perform a classical analysis of the
sectors, and the geometric quantization in the case of $SU(2)$.
The quadratic relation, classically identifying $SU(2)$ as the sphere
$S^3$, is replaced quantum mechanically by a similar condition on
non-commutative operators ('quantum sphere').
The resulting quantum exchange algebra of the chiral group
variables is quartic, not quadratic. The fusion of the sectors leads to
a Hilbert space that is subtly different from the one obtained by a
more direct (un--split) quantization.

\normalsize
\end{quote}
\end{titlepage}
\parindent 5truemm
\parskip 0truemm

\newpage

\section{Introduction}

The theory of Wess--Zumino--Witten (WZW) models has received a lot of
attention since its inception. One of the reasons is that also a quantum
theory corresponding to it \cite{WZW} is well--known, viz. the theory of
unitarizable representations \cite{Kac book} of affine Kac--Moody algebras.
In recent years the canonical structure of the classical model has been
unraveled \cite{BalDomFe,Gaw,PapSpen}. It is fair to say that the
transition between these two was not established by some standard
quantization procedure, like deformation quantisation (DQ) \cite{BFFLS} or
geometric quantization (GQu) \cite{Wood, Snia, tu1}, but rather
guessed. A simplified model, analoguous to WZW models but with a finite
dimensional phase space, has been presented as a toy model for conformal
field theory \cite{AlFad}. This model is based on the free motion of a
particle on a group manifold. The classical theory is straightforward,
and was worked out to some extent in \cite{AlFad} for $SU(2)$, whereas
the proposals of \cite{AlFad} for other groups were derived in
\cite{AlTod}. A quantum theory with the appropriate 'classical limit' was
proposed in \cite{AlFad}, again without systematic derivation.

    In this paper we start from the classical model of a free particle
moving on a compact group manifold. One may take it as an approximation
of the kinematics of the WZW model itself, when one leaves out the
fluctuations and keeps only the zero modes. We will reformulate this
model in such a way that one of the main properties of the WZW model,
viz. chiral splitting, is mimicked. In the WZW model it refers to the
presence of left and right conserved currents, for the particle there are
left and right conserved momenta. It is desirable to use {\it both}
conserved quantities to parametrize the phase space. To this end, we
split the phase space, and the original model arises from a symplectic
reduction. After this splitting, the symplectic structures of both
sectors are identical (up to a sign). Passing to the quantum theory, each
sector can be quantized separately, and afterwards both sectors should be
fused. For the quantization in each sector, we will use the geometric
quantization procedure. In this way the quantum theory emerges as the
result of a systematic procedure. The result after the fusion is close,
but not identical, to the result announced in \cite{AlFad}. For the
chiral sectors separately, it was proposed that the exchange algebra of
the group matrix elements is quadratic, but the relations we find are
quartic instead. Also, we obtain in some sense a quantum sphere: The
classical $S^3$ relation $a^*\, a + b^*\, b=1$ remains valid in the
quantum case, but only with the specific operator ordering given.

    The second section treats the classical model. Its subdivisions treat
subsequently the canonical phase space in the usual form, the
(classically) equivalent viewpoint as a symplectic reduction of a
chirally split extended phase space, the classical treatment of one of
the chiral sectors, and finally, the detailed treatment of $SU(2)$ as an
example and a preparation for the quantum theory. Whereas the classical
treatment can be given for any compact semi--simple Lie group, we
restrict ourselves to $SU(2)$ to carry the quantization program to
completion. This is done in the third section, following geometric
quantization methods. Here we obtain the quartic exchange algebra for the
group elements. In section four, we show how to
fuse the chiral sectors, thus obtaining the quantum version of the
classical reduction. This necessitates the construction of the inverse
matrix of the group element with operator entries ('antipode'). The fifth
section contains some discussion. Finally, in the appendix, we present,
from a slightly different viewpoint, a more detailed exposition of the
structures used in splitting and fusing.

\section{The classical model}
\subsection{Canonical Phase Space}
    We start with a compact group $G$ as configuration space, and let
${\cal G}$ denote the Lie algebra of $G$. The phase space is the tangent
bundle of $G$, with elements $(g, p)$, and the natural choice for the
Lagrangian is the square of the velocity, leading to the equations of
motion
\beq \frac{d}{dt}(g^{-1} \frac{d}{dt} g) = 0. \label{eom} \enq
Defining a momentum corresponding to the trivialization of the bundle
$TG$ by the lift of the left action of $G$ on itself,
\beq p_l= g^{-1} \frac{d}{dt} g, \enq
one sees that it is conserved, since that action is
a symmetry of the system. In terms of this momentum, the symplectic form
is the differential of the following Liouville one--form
\beq \alpha = K(p_l, g^{-1}dg) .\label{L-form} \enq

We can use the conserved momentum  to parametrize the solutions:
\beq
g(t,g_o) = g_o \exp (p_lt) .
\label{left solution}\enq
In the same way, one can also introduce a (conserved) right momentum
$p_r= -  \frac{d}{dt} g  g^{-1} $, and a second
parametrisation of the solution
\beq g(t,g_o) =\exp (-p_rt) g_o  \label{right solution}. \enq

\subsection{Split Phase Space}
Having a large set of global conserved quantities at one's disposal
one naturally wants to use them all.
In the WZW models, in a similar way, conserved left an right currents are
both present. In \cite{BalDomFe}, this was used to parametrize the solutions
(their equations (2) and (3) ) as
\beq
g=P g_0 \bar P
\enq
where $P$ and $\bar P$ are path--ordered integrals of the chiral currents,
and $g_0$ fixes the initial condition.
In this formula, the values of the left and right currents are
not completely independent: this corresponds to the well--known fact
that the representation of left and right zero modes, or the monodromy,
should be the same.
The exponentials should in fact be in the same conjugacy class.
This fact complicates the description of phase space using {\it both}
left and right conserved currents. This feature is reproduced in the
particle model, when we try to use both left and right momenta the particle
phase space. From eqs.(\ref{left solution}) and (\ref{right solution}) it is
clear that, when $p_l$ and $p_r$ parametrize the same solution, they are
in the same conjugacy class (up to a sign)
\beq p_r= - Ad_{g_0} p_l .\enq

Let us try to describe this interdependence. The conjugacy classes
(adjoint orbits) are parametrized by the elements of an arbitrarily chosen
(but fixed) Weyl chamber $W$\label{W choice}.
Now take some values for $p_l$ and $p_r$,
and let $w\in W$ be the element characterizing their orbit. Then, to
reconstruct $p_l$ from $w$, we need in addition some group element
$\sigma_l$ such that $p_l = Ad_{\sigma_l} w$. This group element is
determined up to multiplication (from the right) by the stabilizer of
$w$. For generic (regular) $w$, this stabilizer is isomorphic to the
maximal torus $T$ (fixed by the choice of Weyl chamber) of $G$. So
instead of this group element, it is sufficient to give its right coset,
i.e. an element of $G/T$. To restore $p_r$, we need another independent
element of the coset space, $p_r=-Ad_{\sigma_r} w$. We can conclude that
the space of momenta is
isomorphic to $ W\times G/T\times G/T $. In order to reconstruct $g$, we
have to give some element $h$ of the torus stabilizing $W$. then
$g=\sigma_r h \sigma_l^{-1}$. Consequently for the whole phase space, we
have locally

\beq
TG \buildrel{loc}\over\approx ( W \times G/T\times G/T)\times T.
\label{prepare split}\enq

The details of this construction can be found in the appendix.

    In this picture, the symplectic form induced by \eqn{L-form} on $G/T$
is the symplectic structure of \cite{Kirillov}, while $W$ and $T$ are
canonically conjugated. This suggests that, by doubling the Weyl chamber
$W$ and the torus $T$ in the description \eqn{prepare split}, one can
realize this sytem as a quotient of two symplectic manifolds, each being a
copy of $W\times G$. This is not quite trivial, since one has to show
first that the local description as $G/T\times T$ fits together to $G$
(which is demonstrated in the appendix), and second that a suitable
symplectic reduction can be found. We now describe this product structure
('splitting') and the symplectic reduction ('fusion').

The extended phase space ${\cal P}$ is a product $P_l\times P_r$, where
$P_{l,r}= W_{l,r}\times G_{l,r}$ are two copies the same manifold.
They are equipped with exact symplectic forms, which are exterior
derivatives, $\omega_{l,r}= d\theta_{l,r}$ of the one-forms
\beq
\theta_{l,r}:= K(w_{l,r},g_{l,r}^{-1}dg_{l,r}).
 \label{sectL}
\enq
The symplectic structure on ${\cal P}$ is given as the difference
$\omega_{l}-\omega_{r}$\label{omega dif}.
{}From this expression it is evident that the chiral
sectors are canonically independent, i.e. the functions on the right sector
Poisson-commute with the functions on the left one.
The original phase space $TG$ is restored by symplectic reduction induced
by the (first class) constraints identifying the Weyl chambers,
$w_{l}-w_{r}= 0$.
The  coordinates on the product torus $T_{l}T_{r}$ are canonically
conjugated to these constraints, and consequently the two copies of $T$
are identified as well.
Explicitly, the original variables $g$ and $p$ are given in terms of
the extended phase space variables $\{w_l=w,g_l; w_r=w,g_r\}$ by
\beq
g=g_{l}g^{-1}_{r}, p= Ad_{g_r}w  \label{cfus1}
\enq
This explains the mechanism of this reduction.

\subsection{Classical Chiral Sector}
Here we describe in more detail the structure of a single sector ( and will
omit the subscript). We will express the symplectic form in terms of Lie
algebraic data.

We use the following notation (\cite{Hum}). The set of simple roots
dual to the chosen Weyl chamber (see page~\pageref{W choice} and the
appendix), is $\Delta$, the
set of roots of the Lie algebra is $\Phi$, and the set of positive roots is
$\Phi_+$. The element of the Cartan subalgebra $K$-dual to the root $\beta$
is  $t_{\beta}=i [e_\beta,e_{-\beta}]$. In addition we introduce
$\theta^{\alpha_i}$, the  one--form dual to the simple root $t_{\alpha_i}$
and $\omega^\beta$, the  left invariant one--form dual to the root vector
$e_\beta$. Finally, $w_i$ is the  coordinate in the Weyl chamber in the
basis dual to the one formed by $t_{\alpha_i}$.

The left--invariant one--form $g^{(-1)} dg$ is
\beq
i\sum_{\alpha_i\in\Delta}\theta^{\alpha_i}   t_{\alpha_i} +
\sum_{\beta\in\Phi}\omega^\beta e_{\beta}.
\enq
Taking into account the commutation relations and the orthogonality
properties of the root subspaces in the Lie algebra, one can rewrite the
symplectic form \eqn{omega in K} as
\beq
\omega=\sum_{\alpha_i \in \Delta} dw_i \wedge\theta^{\alpha_i}
+i \sum_{\beta\in\Phi_+} K(w,t_\beta) \omega^\beta\wedge\omega^{-\beta}.
\label{sectsym}
\enq
{}From this, the Poisson brackets follow immediately. The functions on the
group are generated by the matrix elements of  all representations
\cite{Peter-Weil}. Let $M_i (g)(i=1,2)$ be the matrix corresponding to some
chosen representation~'$i$', and $m_i$ the corresponding matrix representation
of the Lie algebra. The  general Poisson brackets of these generators
is given by
\beq
\{M_1,\otimes M_2\} (g)=M_1(g)\otimes M_2(g) r_{12}
\label{exchange relation}
\enq
where one takes the tensor product of matrices,
and the Poisson bracket of the entries.
In \eqn{exchange relation}, we used the exchange operator
\beq
r_{12}= \sum_{\beta\in\Phi_+}\frac{i}{K(w,t_\beta)}
    [m_1(e_{-\beta})\otimes m_2(e_\beta)-
      m_1(e_{\beta})\otimes m_2(e_{-\beta})].
\enq
This formula generalizes the one presented in
\cite{AlTod} (3.24), where it is derived for the specific case where
both '1' and '2' are in the same representation.
The following completes the Poisson bracket relations:
\beq
\{w_i,M\}(g)=M(g) m(t_{\alpha_i}).\label{PB wi M}
\enq
The derivation of the above Poisson brackets becomes easy if one expresses
the exterior derivative of $M$ in terms of the differentiations along the
left invariant vector fields $t_{\alpha_i}^L, e^L_{\alpha}\ \mbox{and}\
e^L_{-\alpha}$ corresponding to the chosen basis elements of ${\cal G}$:
\beq
dM(g)=M(g)\left[ \sum_{\alpha_i\in\Delta}m(t_{\alpha_i}) \theta^{\alpha_i}
 +\sum_{\alpha > 0}m(e_\alpha)\omega^\alpha
 +m(e_{-\alpha)}\omega^{-\alpha}\right].
\enq
This formula follows from the definition of the left invariant vector field
$x^L$ corresponding to $x\in{\cal G}$, $(x^L
f)(g)=\frac{d}{dt}f(ge^{tx})|_0$. In particular, for the matrix function we
have $(x^L M)(g)=M(g) m(x)$. The equation for the hamiltonian vector field
$x^M$ of the function $M$, $i(x^M) \omega+dM=0$, takes a very simple form
if $x^M$ is expanded in terms of the left invariant basis. The solution is
given by
\beq
x^M(g)=M(g)\left[\sum_{\alpha > 0} \frac{i}{K(w,t_\alpha)}
[m(e_{-\alpha})e^L_\alpha -m(e_{\alpha})e^L_{-\alpha}]
-\sum_{\alpha_i\in\Delta}m(t_{\alpha_i})\frac{\partial}{\partial w^i}\right]
\enq
and eqs. (\ref{exchange relation},\ref{PB wi M}) follow from the definition
of the Poisson bracket.

As demonstrated, the classical treatment can be given quite generally.
For the quantisation, we will not be able to work it out completely,
but will restrict ourselves to $SU(2)$. For that reason, and to provide an
explicit example for the above, we treat this case in detail in the next
subsection.

\subsection{The    $SU(2)$ case}
We shall identify $TSU(2)$ with $S^{3} \times su(2)$ , where $su(2)$ is the
space of $2 \times 2$ antihermitian complex matrices.
 This space is spanned by $l_i := {i \over 2} \sigma _i$, where
$\sigma _i$ are the Pauli matrices. The invariant form $K$ is simply
proportional to a trace of a product:
\beq
K(p,\tilde p ) := -2 Tr(p\tilde p ) .
\label{42}
\enq
 Now we parametrize the space of orbits.
The Weyl chamber we choose is
\beq
W:= \{\tilde w:\tilde w = wl_3; w>0 \}
\label{43a}
\enq
(temporarily denoting the Weyl chamber element by$\tilde w$ to avoid
confusion with the real value $w$).
The remaining
information about momenta is contained in local sections $\sigma
$ (\ref{secsigma}) of $G \rightarrow G/T$.
Take $p$ and $\tilde p$ in the same conjugacy class.
Then there  exist sections $\sigma, \tilde\sigma$ such that
\beq
p = Ad_{\sigma} wl_{3} , \qquad \tilde{p} = Ad_{\tilde\sigma} wl_{3}.
\label{45}\enq
In accord with the approach presented in the previous section we will
now focus our attention on the left sector only, and we will for convenience
call this sector $(P,\omega)$.

 Take the stereographic  parametrization of $G/T \simeq S^2$ .
It is defined by two neighbourhoods $U_\pm$, which are identified with two
complex planes with the transition
$z_+ = z_-^{-1}$ on $U_{+-}:= U_+\cap  U_-$.
Take  two local sections $\sigma_\pm : S^2 \rightarrow SU(2)$:
\begin{equation}
\sigma_+ = f_+ \left[\matrix{1&iz_+\cr i\bar z_+&1\cr}\right] \quad ,
\quad
\sigma_- = f_- \left[\matrix{-i\bar z_- & -1\cr1&iz_- }\right] \quad ,
\label{49b}
\enq
where $f_{\pm}=f(z_\pm),f(z) := (1 + z \bar z)^{-1/2}$.
These cover the two neighbourhoods
\beq
        O_\pm := \{p: w\pm p_3 \neq 0 \} \subset \beta^{-1} (wl_3);
        \quad r \in {\bf R}_+
\enq
of the adjoint orbit by $ \sigma _{\pm }wl_{3}\sigma ^{-1}_{\pm }= p$.
Any group element $g$ can be represented locally as
\beq
g_{\pm } = \sigma _{\pm } e^{i\sigma_3q_\pm} ,
\label{52}\enq
and on $U_{+-}$ we  have
\beq
\sigma _{-+} e^{iq_+ \sigma_{3}} =e^{iq_- \sigma_{3} } \quad , \quad
\sigma _{-+}=
 \mid z_+\mid ^{-1}\left[\matrix{i\bar z_+&0\cr0&-iz_+}\right] \quad ,
\label{transition}
\enq
and consequently
\beq
g_{+} = \sigma _{+} e^{iq_{+}\sigma_{3}} = \sigma _{-}
e^{iq_{-}\sigma_{3}} = g_{-}  ,  \enq
so that the group element is well defined globally.
Having displayed the above transition maps once, we will drop the $(\pm )$
indices, but emphasize that our  calculations are globally true.

We represent an element
$g\in SU(2)$ by a $2\times 2$ matrix:
\beq
g \equiv  \left[\matrix{ a &-b^{*}\cr b & a^{*}}\right]
,\qquad a^{*}a+b^{*}b = 1 .
\label{55}
\enq
{}From (\ref{52}) it follows that $a$ and $b$ have the following local
coordinate expressions :
\beq
a = fe^{iq} ;\qquad  b = i\bar z fe^{iq}
\label{56}
\enq
The ('left') Louville form $\theta  =
K(\tilde w,g^{-1}dg) $, where $\tilde w=wl_{3}$  ,
is locally equal to:
\beq
\theta = w dq + {iw\over 2} f^{2}(\bar z dz - zd\bar z ) ,
\label{57}
\enq
and the symplectic form reads:
\beq
\omega  := d\theta  = dw\wedge dq + {i\over 2} f^{2}dw\wedge (\bar z dz-
zd\bar z) + iwf^{4}(d\bar z\wedge dz).
\label{58}
\enq
 Having a local expression for $\omega $ we can find (local)  expressions
for the (global) hamiltonian vector fields
\begin{eqnarray}
X_{a} &=&e^{iq}( {1\over 2w} \mid z\mid ^{2}f \partial _q -
if \partial _w + i z {1\over fw} \partial _z )
\nonumber\\
X_{b} &=& e^{iq}(- {i\over 2w} \bar z  f \partial _{q} +
\bar z f \partial _w + {1\over fw} \partial _z )
\nonumber\\
X_{w} &=& \partial _{q}
\label{vector fields}
\end{eqnarray}
 and the Poisson brackets $\{f,g\}:=X_{f}g$ :
\begin{eqnarray}
\{ a^{*}, a \} = {i\over w} b b^{*} &,&\quad \{\, a , b \,\} = 0\qquad ,
\nonumber\\
\{ a , b^{*}\} = {i\over w} a b^{*} &,&\quad
  \{ b , b^{*}\} = - {i\over w}a a^{*} ,
\label{Poisson a b} \\
\{\, a , w\, \} = - i a \ &,&\quad \{ a^{*}, w \} = i a^{*} ,
\nonumber\\
\{\, b , w \,\} = - i b \ &,&\quad \{ b^{*}, w \} = i b^{*} ,
\label{Poisson w}
\end{eqnarray}

There is an obvious generalization of the above algebra.
Recall from the previous section that the definitions of Liouville and
symplectic forms (\eqn{omega in K}) depend on an identification of the space
of orbits with a fixed Weyl chamber. We are free  to
choose  this Weyl  chamber in an arbitrary way. If we take $Ad_{g_o} l_3$
instead of $l_3$ in \eqn{theta},
 the Liouville form $\theta $ is
\beq
\theta  = K(wg_ol_3 g^{-1}_o,\omega ) ,
\label{newtheta}
\enq
 where $g_o \equiv  \left[\matrix{\alpha &-\beta^{*}\cr
\beta & \alpha^{*}}\right]$ is an arbitrary element of $SU(2)$.
 The symplectic form can be computed from  $\omega  := d\theta $
 and the resulting Poisson  algebra in general depends on $g_o$:
\begin{eqnarray}
\{a,a^{*}\} &=&
- {i\over w} c_{1}bb^{*} - {i\over 2w} (ab\bar{c}_{2} + a^{*}b^{*}c_{2})
\nonumber\\
\{\,b,b^{*}\} &=& - {i\over w} c_{1}aa^{*} + {i\over 2w} (ab\bar{c}_{2} +
        a^{*}b^{*}c_{2})
\nonumber\\
\{a,b^{*}\} &=&
{i\over w} c_{1}ab^{*} + {i\over 2w} (aa\bar{c}_{2} - b^{*}b^{*}c_{2})
\nonumber\\
\{\,a,b\,\} &=& {i\over 2w} c_{2}
\label{new Poisson a b}\\
\{\,w,a\,\} &=& ic_{1}a - ic_{2}b^{*}
\nonumber\\
\{\,w,b\,\} &=&  ic_{1}b + ic_{2}a
\label{new Poisson w}
\end{eqnarray}
where
\beq
c_{1} := \mid \alpha\mid ^{2} - \mid \beta\mid ^{2} ;\qquad c_{2} :=
2\alpha^{*}\beta;\qquad c^{2}_{1} + \mid c_{2}\mid ^{2} = 1 .
\label{63}\enq

This is a generalization of the algebra of eqs.(\ref{Poisson a b}) and
(\ref{Poisson w}), to which it is obviously equivalent. The former
algebra is preferable to carry through the quantisation
in the next section, because it is the only one for which the
functions $a$ and $b$ on the group (Poisson) commute.

One can easily check that the change of Weyl chamber described in
\eqn{newtheta}) is equivalent to the right action of $g_{0}$ on $G$.
It is natural to ask whether there exists a Poisson structure on $G$
such that the action
\beq
G \times P \ni (g_{0},(w,g)) \rightarrow (w,gg^{-1}_{0}) \in P
\label{64}\enq
preserves the relations of eqs.(\ref{Poisson a b}-\ref{Poisson w}).
 The answer is affirmative for the quadratic relations \eqn{Poisson a b},
if one imposes the following Poisson relations on the group parameters
$\alpha$ and $\beta$:
\begin{eqnarray}
\{\, \alpha  , \beta \,  \} &=&- {i\over w} \alpha \beta \,\quad ;\quad
\{ \alpha ^{*}, \beta ^{*}\} = {i\over w} \alpha ^{*}\beta ^{*} ;
\nonumber\\
\{ \alpha  , \beta ^{*}\} &=& - {i\over w} \alpha \beta ^{*} \quad ; \quad
\{\, \alpha  , \alpha ^{*}\} = {2i\over w} \beta ^{*}\beta  ;
\nonumber\\
\{ \beta ^{*}, \beta  \} &=& 0\quad .
\label{Woron}
\end{eqnarray}
These brackets define a one--parameter family of $SU(2)$ Lie--Poisson groups,
which was studied in \cite{LuWein}.
It can be identified as a one-parameter family of classical versions
 of the algebra of Woronowicz \cite{wor,jeu}.
As far as the commutation rules with $w$ are concerned, \eqn{Poisson w},
the above action of the Lie--Poisson group \eqn{Woron} does {\it not}
preserve the Poisson brackets of $w$ and $a,b$.

We close the classical treatment of $SU(2)$ with a remark.
We may think about the commutative ring generated by
$a,b,a^*,b^*$ and the single relation $aa^*+bb^* = 1$ as the algebra
of functions on the manifold of SU(2). It is not usefull to think about
this manifold as  a group equipped with a multiplication. The structure that
is actually used, and that is compatible with the symplectic structure,
is that of a manifold with a simply transitive action of the group on it.
Only the left action of $SU(2)$ on the  manifold is consistent with the
Poisson algebra of the functions $a,b,a^*,b^*$ and the Weyl chamber
variable $w$.

\section{Quantum Chiral Sector}

Generally speaking, one may apply a variety of methods to quantize a given
classical system. Among the rather unambiguous methods, we mention
deformation quantization \cite{BFFLS,reif} and geometric
quantization (GQu) \cite{Wood,Snia}.
We will use the second method (the simplest version \cite{geomas,Snia} is
sufficient), because it gives not only the quantum algebra of operators,
but also the Hilbert space representation.

 First we have to build a complex line bundle over the phase space with
a connection, the curvature of which is proportional to the symplectic form
$\omega$. This step is very simple because $\omega$ is exact, and
consequently the line bundle is trivial.
We can take a global section $\lambda_{0}$ and define the connection by
introducing the following covariant differential operator
\beq
D\lambda _o = - {i\over \hbar } \theta \otimes \lambda _o
\label{68}
\enq
 where $\hbar $ is the Planck constant divided by $2\pi $ and $\theta$ is
the Liouville form.
Since $\theta $ is real, we can normalize the section by the condition
\beq
H(\lambda _o,\lambda _o) = 1 ,
\label{69}
\enq
 which uniquely defines the hermitian form $H$ on the line bundle
and fixes the scalar product of the sections (pre-quantum wave-functions
\cite{Wood,Snia}).

 As is well--known, the pre-quantum wave-functions depend on all phase space
variables and consequently they do not yet give an adequate quantum
description of the physical system:
it is neccessary to find the space of wave-functions depending on,
roughly speaking, the spectrum of a maximal algebra of commuting observables.
 Technically this is achieved by choosing a polarization, i.e. an involutive
lagrangian distribution $F\subset TP$, and imposing the condition on the
sections to be covariantly constant along $F$.

In our case we want to represent the algebra of functions on the group
manifold $G$. Then it is natural to take $F$ to be spanned by the two
vector fields corresponding to the Poisson commuting functions
$a$ and $b$\footnote{The other choice is to take $a^*$ and $b^*$, which for
a single sector amounts to a change in sign of $w$. The choice in the other
sector (see further) then should be adapted to this.}:
\beq
F :=\hbox{ span}\{ X_a\,,\, X_{b} \} .
\label{70}
\enq
This way, the variable $w$ turns out not to be independent, but the
corresponding operator will be a function of $a,b,a^*,b^*$.
The non-commutative algebra formed by the operators $a,b$ and their
conjugates (which as we shall see will generate the spectrum of the system)
will  correspond to the "algebra of functions" on a 'quantum' $S^3$
manifold, but not on a quantum $SU(2)$ group in the sense of Hopf algebras.

With this polarization, the wave functions
\beq
        \Psi = \varphi \lambda_o
\enq
have to satisfy the conditions
\beq
        \nabla _{X_{a,b}} \Psi =0
\enq
which are equivalent to
\beq
{\partial \varphi \over \partial w} = 0\qquad\hbox{   and }\qquad
\Big ( {\partial \over \partial z} - {i\over 2} (1+z\bar z)^{-1}\bar z
{\partial \over \partial q} \Big ) \varphi  = 0.
\label{74}
\enq
They are satisfied by any superposition of the following fundamental
solutions
\beq
\varphi _{k} = (1+z\bar z)^{-k/2}e^{ikq}h(\bar z)\qquad
(\hbox{no sum over }k) ,
\label{75}\enq
where $h$ is a holomorphic function of $\bar z $.
The sections are then given by
\beq
        s_{k} = \varphi _{k}\otimes \lambda _o.
\label{76}
\enq
To normalize these wave functions, one can not use
the symplectic scalar product defined by the density ${\mid \omega\mid}^{2}$,
since it includes integration of $w$-independent functions along
the non-compact Weyl chamber.
Within the framework of geometric quantization, there are two
methods of dealing with this difficulty. The first multiplies
the wave-functions by  half-densities \cite{Snia,tu23}. the second one
uses half-forms \cite{geomas,Snia}.
In our case, after application of the half-form method, one obtains
the following representation for the scalar product:
\beq
(s,s')=\int \limits_{S^3} (1+z\bar z)^{-2} \bar{\varphi } \varphi '
\mid d\bar z \wedge dz\wedge dq\mid .
\label{77}\enq
Following the standard procedure of GQu, the operators
corresponding to the classical functions $a$ and $b$ are (by our choice
of polarisation) simply multiplication operators:
\begin{eqnarray}
        a\quad\rightarrow \quad \hat{a} &=& \hbox{mult  }a\nonumber\\
        b\quad\rightarrow \quad \hat{b} &=& \hbox{mult  }b \ ,
\end{eqnarray}
whereas
\beq
w\quad\rightarrow \quad \hat{w} =
\hbar(-i {\partial\over \partial q} + 1).
\enq

Now let us analyse the detailed form of the wave-functions \eqn{76}. Let
$s_{k} = (1+z\bar z)^{-k/2}e^{ikq}h\lambda _o$  , where $k\in {\bf Z}$
has some fixed  value  and $h = h(\bar z )$   is   holomorphic.
Let us transform $s_{k}$ from the neighbourhood $U_-$ to
$U_+$ (see \eqn{transition}):
\begin{eqnarray}
s_{k-}(q_-,z_-,\bar z_-) &=& (1+z_-\bar z_-)^{-k/2}\exp(ikq_-)
h(\bar z_-f )\lambda _o =
\nonumber\\
&=& \mid z_+\mid ^{k}(1+z_+\bar z_+)^{-k/2}\exp (ikq_+)
\big ( { \bar z_+
\over \mid z_+\mid } \big)^{k} h(\bar z ^{-1}_+)\lambda _o
\nonumber\\
&=& (1+z_+\bar z_+)^{-k/2}\exp (ikq)_+) \bar z ^k_+
h (\bar z ^{-1}_+) \lambda _o .
\end{eqnarray}
 Clearly in order to keep the holomorphic property of the section $s_k$
one should restrict $h$ to be a polynomial of degree not higher than $k$
and $k$ has to be non-negative. This argument is of a
purely geometrical nature. One can arrive at the same conclusion
for the degree of $h$ using a normalisation condition: the sections \eqn{76}
have finite norm with respect to the scalar product \eqn{77} if and only
if the above condition on $h$ is satisfied.
 Consequently the vectors
\beq
s_{k,j}:= e^{ikq}(\bar z)^{j}(1+z\bar z)^{-k/2}\lambda _o \quad;
\quad j \le k
\label{80}
\enq
 form a basis in the Hilbert space of quantum states.
For the scalar product of the basis elements we get:
\begin{eqnarray}
(s_{k,n}, s_{m,l}) &=& \int e^{(m-k)q}(z)^n
(\bar z )^{l}(1+z\bar z)^{-(m+k+4)/2}
\mid d\bar z dzdq\mid
\nonumber\\
&=&(2\pi )^{2} \delta _{m,k}\delta _{j,l}
\big[\big( ^k_j\big )(k+1)\big]^{-1}.
\label{scalar product}
\end{eqnarray}
We normalize the states and relabel them according to the rule:
\beq
        s_{k,n} \Rightarrow \psi_{j,j_3}
\enq
where
\beq
        j = {k\over 2} \quad;\quad j_3 = {k \over 2} - n .
\enq
The values of $j$ are then non-negative half-integers and for fixed $j$
the value of $j_3$ ranges from $-j$ to $j$ step 1.

The action on these states by the quantum
operators corresponding to the generators of the algebra of functions on
$S^3$ is then given as (here and in the sequel we identify $a$ with
$\hat{a}$ etc):
\begin{eqnarray}
a \psi_{j,j_3} &=& \Big( {j+j_3+1 \over 2j+2}
\Big)^{1/2} \psi_{j+{1 \over 2},j_3+{1 \over 2}}
\nonumber\\
a^{*}\psi_{j,j_3} &=& \Big( {j+j_3\over 2j+1}
\Big)^{1/2} \psi_{j-{1 \over 2},j_3-{1 \over 2}}
\nonumber\\
b \psi_{j,j_3} &=& \Big( {j-j_3+1\over 2j+2}
\Big)^{1/2} \psi_{j+{1 \over 2},j_3-{1 \over 2}}
\nonumber\\
b^{*}\psi_{j,j_3} &=& \Big( {j-j_3\over 2j+1}
\Big)^{1/2} \psi_{j-{1 \over 2},j_3+{1 \over 2}} .
\label{82}
\end{eqnarray}
The $a^*$ operators can most easily be obtained from the fact that the
quantization method, including the scalar product \eqn{scalar product},
guarantees that it is the hermitian conjugate of $a$. The other method,
using a varying polarization, is much more involved.

Notice that $\Psi = \psi_{0,0}$ is the unique state anihilated by
both $a^*$ and $b^*$.
The whole Hilbert space of states can be generated by the successive
action of $a,b,a^*,b^*$ operators, hence we will call their algebra
the spectrum generating algebra (SGA).
We can find explicitly  the relations satisfied by the
generators of this algebra.
First of all, the condition that shows the classical $SU(2)$ group
to be a sphere, reappears in the form
\beq
a^{*}a + b^{*}b = 1 ,
\label{sphere}
\enq
with the ordering of the operators  fixed. With the opposite ordering,
the left hand side differs from $1$ by a term of order $\hbar$, see
\eqn{w in SGA}. Hence the sphere has become a 'quantum sphere'.

Secondly, corresponding to the classical \eqn{Poisson a b},
\begin{eqnarray}
 [\, a,\, b\,]   &=&   0     ,              \nonumber\\
{ [ a, a^*]} &=&\ (aa^* + bb^* -1) b^*b, \nonumber\\
{ [\, b, b^*]} &=&\ (aa^* + bb^* -1) a^*a, \nonumber\\
{ [ a,\, b^*]} &=& -(aa^* + bb^* -1) b^*a .  \label{quantalgebra}
\end{eqnarray}
These quartic relations (and similar ones with the bracket on the right
hand side at the end ) could be written in a pseudo-quadratic form,
like the classical ones, \eqn{Poisson a b}, using $w$, which parametrizes
the Weyl chamber, and which is an independent variable classically.
Here however,  $w$ is not an independent operator,
\beq
w= -\hbar (aa^* + bb^*-1)^{-1}.  \label{w in SGA}
\enq
 It has a discrete spectrum given by
\beq
\{ \hbar (2j+1),\qquad j \in {\mid {\bf Z}\mid \over 2} \} .\label{discr w}
\enq
It commutes with $a a^*$ and $b b^*$.
Although the
quantum system is still living on a 'quantum group-manifold', \eqn{sphere},
the relations \eqn{quantalgebra} do not define a {\it quantum group }
structure. This was already visible at
the classical level, as the Poisson structure was not a Lie-Poisson
structure \cite{LuWein}, and moreover the right group action does not preserve
the symplectic structure, so one is not able to define the group
multiplication in a compatible way.

 The quantum hamiltonian of a free particle can be  represented in terms
 of the generators of the SGA:
\beq
        H = -{\hbar^2 \over 2}  {{(aa^* + bb^*-2)(aa^* + bb^*)}\over
                {(aa^* + bb^* -1)^2}}
\label{92}
\enq
and its eigenstates, with eigenvalues equal to
${1 \over 2} j(j+1)\hbar^2$, are  $\psi_{j,j_3}$.
 Consequently the time evolution is diagonalised.
The operators corresponding to the classical
left momenta \eqn{15} are the following:
\begin{eqnarray}
        J_3 &=& {\hbar \over 2} {{a a^*-b b^*}\over {a a^*+b b^*-1}}
\nonumber\\
        J_+ &=& \hbar {{ab^*} \over {aa^*+bb^*-1}}\  =\ J_-^*
\end{eqnarray}
and their action on the states is standard:
\begin{eqnarray}
J_3 \psi_{j,j_3} &=& \hbar j_3 \psi_{j,j_3}
        \nonumber\\
J_\pm \psi_{j,j_3} &=& \hbar \big ( (j \mp j_3 )(j\pm j_3+1)
        \big )^{1/2}\psi_{j,j_3\pm 1} .
        \label{JJ3}
\end{eqnarray}
Comparing \eqn{JJ3} with \eqn{82} one can see
that the operators $a$ and $b$ together with their conjugated form
a kind of 'square roots' of the Lie algebra formed by $J_3,J_\pm$.

In conclusion, the Hilbert space generated by the SGA out of
the vacuum $\Psi$ is  a direct
sum of all irreducible unitary representations (with multiplicity 1) of the
group $SU(2)$. Symbolically
\beq
        SGA\ \Psi = {\cal H} = \bigoplus_{j \in {\mid {\bf Z}\mid \over 2}}
        {\cal H}_{j}
\enq

This ends the application of the geometric quantisation method to the left
sector of our extended phase space.
The procedure can be repeated for the right sector.
The quantum description is obtained by exchanging
the roles of $a$ with $a^*$ and $b$ with $b^*$. This is not a matter of
choice, but is imposed by  the
difference in sign of the symplectic forms of the right and left sector,
see \eqn{symplectic product} and page~\pageref{omega dif}

\section{Fusion}
The classical fusion \eqn{cfus1} was  straightforward, doing a symplectic
reduction.
Now we investigate the quantum version of this  reduction.
The first step is to form the quantum analog of
the symplectic product of the classical sectors.
Since at the classical level these sectors are completely separated, i.e.
the observables of different sectors Poisson-comute, it is natural to
take  the tensor product of Hilbert spaces as the Hilbert space
for the quantum product:
\beq
        {\bf H} = {\cal H}_{l}\bigotimes {\cal H}_{r}\ .
\label{Hr}
\enq
The reduction should now identify the Weyl chamber elements of the left
and right sectors, which means that we should extract from this space
the kernel of the quantum constraint $w_l-w_r=0$. Since $w$ had a discrete
spectrum, \eqn{discr w}, this is quite trivial:
\beq
        {\bf H}_o = \bigoplus_{j \in {\mid {\bf Z}\mid \over 2}}
        {\cal H}_j \otimes {\cal H}_j \ .
        \label{Ho}
\enq
This space is isomorphic with
${\cal L}^{2} \big(SU(2) , d\nu \big)$ , where $\nu$ is an invariant
measure. The SGA of this space is simply the commutative algebra of the
matrix elements of all the unitary representations of $SU(2)$
\cite{Peter-Weil}.
Therefore one should be able to recover the generators of this algebra
out of the generators of SGAs of the left and right sector.
At the classical level the prescription is given by the formula
\eqn{cfus} of symplectic reduction. To extend it to the
quantum level we need the antipode - the analog of the group inverse
for the algebra \eqn{quantalgebra}.
Let us introduce matrices with operator entries:
\beq
        (g_{i,j}^r) =   \left[ \matrix{ a_r & -b_r^*\cr b_r & a_r^*}\right]
\enq
and let
\beq
        (S(g^r)_{i,j}) = \left[ \matrix{ a_r^*(1+{\hbar \over w_r}) &
        b_r^*(1 + {\hbar \over w_r}) \cr -b_r & a_r} \right]
\enq
where $w_r = -\hbar (a_r^*a_r + b^*_rb_r -1 )^{-1}$. It can be verified
that
\beq
 S(g^r)_{i,j} g^r_{j,k} = g^r_{i,j} S(g^r)_{j,k} = \delta_{i,k}
\label{antipode}
\enq
which means that $S$ is the desired antipode (although not in the sense
of the Hopf algebras). Of course there is a similar operation in the left
sector. The generators of the SGA for the fused model
are the entries of the following 'coproduct' matrix (compare \eqn{cfus1}):
\beq
        G_{i,j} := g^l_{i,k} \otimes S(g^r)_{k,j}
        \equiv  \left[ \matrix{ A & C \cr B & D}\right]
\label{coproduct G}
\enq
Explicitly, they are given by
\begin{eqnarray}
        A &=& a_l\otimes a^*_r (1+{\hbar \over w_r}) + b^*_l \otimes b_r
\nonumber\\
        B &=& b_l\otimes a^*_r (1+{\hbar \over w_r}) - a^*_l \otimes b_r
\nonumber\\
        C &=&-b^*_l \otimes a_r+a_l\otimes b^*_r (1+{\hbar \over w_r})
\nonumber\\
        D &=&\ a^*_l \otimes a_r + b_l\otimes b^*_r (1+{\hbar \over w_r}) .
\label{G matrix}
\end{eqnarray}
They form a commutative algebra on the kernel of the constraint operator,
as can be checked by straightforward but rather tedious calculations. The
determinant of this matrix is equal to 1 and there are no other quadratic
relations between the matrix elements. Note that this coproduct matrix
is not the operator that one obtains by quantising the classical
matrix elements, obtained by putting $\hbar=0$ in \eqn{G matrix}.
Instead, it was directly constructed out of the quantum operators, with the
intention to obtain a commuting set.

    We now make a comparison between the results of the present split
quantization (SQ) and the perhaps more straightforward one, which consists
in direct quantization (DQ) in the cotangent bundle, without decomposing
it in left and right sectors first. The classical theories are of course
completely the same, see the reduction theorem in the appendix.
Both quantizations give rise to a $2\times 2$ matrix of
commuting functions. Also, both matrices have determinant equal to 1. What
is different however, is that DQ also automatically keeps the unitarity of
this matrix, whereas SQ violates it, as can be seen easily from
the formulas for $G_{i,j}$. This is the price paid for the fact that
the antipode, \eqn{antipode}, is not related to $g$ by
matrix conjugation. We recall that its form was instead designed
to obtain the matrix of commuting quantities $G$, \eqn{coproduct G}.
If one wishes to restore the unitarity
one has to modify the scalar product in ${\bf H}_o$.
This can be done by redefining the scalar product in one
of the sectors, or both, for example:
\beq
        \lan \psi , \psi' \ran := ( \psi , {1 \over w_r} \psi')_{old}
\enq
The conjugation with respect to the new product (denoted by ${\ }^\dagger$)
is related  to the old one (which was denoted by ${\ }^*$):
\begin{eqnarray}
        a_r^\dagger &=& a_r^*(1+{\hbar \over w_r})\nonumber\\
        b^\dagger_r &=& b^*_r(1+{\hbar \over w_r})
\end{eqnarray}
The tensor product (\ref{Hr}) and also the Hilbert space (\ref{Ho}) are then
equipped with a pairing, with respect to which $D$ is
conjugated to $A$ and $-C$ is conjugated to $B$. The unimodularity
property becomes then equivalent to unitarity.
The new scalar product is weaker than the old one so the new Hilbert space
contains more vectors in the completion of the set of states with fixed
spin. Since $w$ commutes with the hamiltonian, and also with the spin
operators, transition probabilities are not influenced by
this change. This might conceivably be different if this model is used as an
ingredient for an interacting theory.

\section{Conclusion and outlook}

    In this paper we have shown how the classical quadratic Poisson
algebra of the chiral sectors of a particle model can be quantized by
applying a systematic procedure, geometric quantization. It gives not
only `quantum deformed` algebraic relations, but also the representation
of the corresponding operators in a Hilbert space. We carried out this
procedure very explicitly for $SU(2) \sim S^3$, obtaining a quantum
version of the sphere relation, and a quantum version of the inverse. The
exchange relations, written in terms of independent variables, are
quartic. We were not able to push through the quantization procedure in
the general case of an arbitrary compact Lie group. For $SU(2)$ we could
easily specify the polarization by the Poisson--commuting functions $a$
and $b$, but the generalization to arbitrary groups is not obvious. It is
of course reasonable to expect that in the general case, the Hilbert
space of states of the chiral sector will be a direct sum of all
irrreducible representation modules of corresponding Lie algebra. For the
chiral sectors, it is not clear that the exchange algebra of the
quantized matrix elements will be quartic in general. It might well be
that the degree of these relations is correlated with the rank of $\cal G$.

We may add that the geometrical mechanism of the chiral splitting
of the canonical WZW theory is exactly the same as in the case of free
point particle treated in detail in this paper. The (affine) orbits of the
loop group describing the
conjugacy classes of chiral momenta are also parametrized by (single)
Weyl chamber. In fact the chiral sector of WZW is composed of a point
particle and fluctuation modes, described by the loops based at unity
\cite{bham}. Both are coupled, and this coupling can be seen to underlie
a deformation of the classical exchange algebra \cite{BlokFad} of the zero
modes of WZW. The point particle exchange algebra can be recovered after
shrinking the loop to a point \cite{Zbig later}.

\medskip
{\bf Acknowledgements }

The authors would like to thank to Prof. K.Gaw\c{e}dzki for a discussion and
helpful remarks.
\renewcommand{\thesection}{}
\section{Appendix}
In this appendix, we present a point of view on the classical model that is
slightly different from the one in the text. At the same time, we provide a
more detailed description of splitting and fusion, with proofs of the main
statements.

To  describe the system we trivialize the bundle $TG$
by means of the left (right) action of $G$ on itself:
\beq
L_{g_o}g := g_o g \in  G \quad ;\quad R_{g_o}g := gg_o^{-1} \in  G
\quad; \forall g_o \in  G .
\label{actions}
\enq
Both are lifted canonically  to actions on $TG$.
We shall use the one corresponding to the left action to identify
globally $TG$ with $G \times {\cal G}$.

In these trivializing coordinates each element of $TG$ is represented
by a pair $(g,p)$ and the Lagrangian is given by:
\beq
L(g,p) = \frac{1}{2}K(p,p)
\label{6}
\enq
where this  $K$ is the $Ad$-invariant form on ${\cal G}$.

Let us now pass to the Hamiltonian analysis of the system.
By means of the $Ad$-invariant form $K$ we identify ${\cal G}$ with
${\cal G}^*$. The canonical Liouville one--form is then given by
\beq
\alpha  = K(p,g^{-1}dg) .\label{ap:L-form}
\enq
The symplectic form $\Omega $ is the exterior differential of $\alpha $:
\begin{equation}
\Omega  = K(dp,g^{-1}dg) - K(p,g^{-1}dg\wedge g^{-1}dg).
\end{equation}
Together with the hamiltonian
\beq
H = {1 \over 2}K(p,p) ,
\label{ham}
\enq
this structure describes the dynamics (kinematics) of a free particle on
the group $G$.

The information about the global constants of motion is very well encoded in
terms of momentum mappings \cite{GuillStern} corresponding to the
group actions \eqn{actions}.
In the coordinates on $G \times {\cal G}$ the lifts of the actions
\eqn{actions} are
\beq
\l _{g_o}(g,p) = (g_o g,p) \quad;\quad r_{g_o }(g,p) = (gg^{-1}_o,Ad_{g_o}p).
\label{14}
\enq
They are symplectic with respect to $\Omega $
and their Hamiltonian realizations are given
by the following momentum mappings:
\beq
J_l (g,p) := -Ad_g p \quad ;\quad J_r (g,p) := p.
\label{15}
\enq
Since the Hamiltonian is $Ad$-invariant it is obvious that
it Poisson-commutes with the momenta \eqn{15} (i.e. they are conserved).
The mapping $I$ :
\beq
        TG\ni (g,p)\mapsto I(g,p):=(-J_{l}(g,p),J_{r}(g,p))=(Ad_{g}p,p)
        \in {\cal G}^{2}
        \label{mapI}
\enq
 projects the phase space onto the set of constants of motion.

We parametrize  the  image  $I(TG)$ in a special way in two steps.
The first step is quite standard, whereas the second step is specific to
our purposes.
\begin{itemize}
\item
Let the map $\beta: {\cal G} \ni  p \rightarrow  \beta (p) \in  {\cal W}$
be  the  projection  onto  the space of adjoint orbits.
This  mapping  can be defined  by  all  independent
Casimir polynomials. The number $n$ of independent Casimirs (the rank of $G$)
is equal  to  the  dimension  of  the  target  for $\beta $.
The mapping $\beta $ restricted to  the  open  subset ${\cal G}^o$  of
regular points defines a smooth fibration of ${\cal G}$ over ${\cal W}$ with
the adjoint orbit ${\cal O}$ of maximal dimension as a fiber.
The space ${\cal W}^o$ of
the  orbits  of maximal dimension is an open, convex subset of
${\bf R}^{n}$  and consequently the fibration $\beta $ is trivial.
Therefore we have :
\beq
{\cal O} \hookrightarrow  {\cal G}^o \approx  {\cal W}^o\times {\cal O}
\buildrel{\beta }\over\longrightarrow {\cal W}^o .
\label{18}
\enq
\item
Define the fibered product:
\beq
\Delta _{\beta } := \{\rule{0mm}{8mm}(\tilde{p},p) \in  {\cal G}^{2} ;
\beta (\tilde{p}) =
\beta (p)\} = {\cal G}\times_\beta {\cal G}
\label{19}\enq
 with the projection being given by
\beq
\Delta _{\beta } \ni  (\tilde{p},p) \mapsto  \beta (\tilde{p},p)
\equiv \beta (p) \in  {\cal W}\ .
\label{20}\enq
 For $\Delta ^o_{\beta } := \beta ^{-1}({\cal W}^o)$ we can write
\beq
\Delta ^o_{\beta } \approx  {\cal W}^o\times {\cal O}\times {\cal O} .
\label{21}\enq
\end{itemize}
It is easily seen that the image of $I$, \eqn{mapI}, is
$I(TG) = \Delta _{\beta }$.
 Defining $\quad TG^{0} := I^{-1}(\Delta ^{0}_{\beta })\quad$
we now show:

\begin{quote}
{\it
\beq
TG^o \buildrel{I}\over\longrightarrow \Delta ^o_{\beta }
\label{22}
\enq
 is a principal fiber bundle with the maximal torus $T$  as  typical fiber.}
 \end{quote}
To show this, we begin with some observations  about the structure of
$\Delta ^o_{\beta }$.
We identify ${\cal W}^o$ with some Weyl chamber $W \subset {\cal G}$.
The interior of $W$
(denoted by $W^o$) can be identified with the space of regular orbits.
Let $T$ be the maximal torus stabilizing $W^o$.
Now we define local sections
\beq
\tilde \sigma \times \sigma :\tilde  U \times U \rightarrow G
\label{secsigma}
\enq
where $\tilde U \times U \subset G/T \times G/T$ is a local neighbourhood,
and let
\beq
p = Ad_{\sigma } w ; \tilde p = Ad_{\tilde\sigma } w .
\label{24}\enq
be a pair of elements of the orbits corresponding to $w\in W$.
 Then for the bundle neighbourhood
\beq
V := I^{-1}( W^o \times \tilde U\times U)
\enq
we have
\beq
V = \{(g,p); \tilde p = Ad_g p, (\tilde p,p) \in W^o\times \tilde U\times U\}
\label{25}\enq
 and the trivializing map:
\beq
V \ni  (g,p) \rightarrow  \psi (g,p):=((\tilde p ,p),\tilde \sigma ^{-1}
(\tilde p)
g\sigma (p) )\in  ( W^o\times \tilde U \times U)\times T ,
\label{26}\enq
 where the identification  of $(W^o\times \tilde U\times U)$  with  a
corresponding neighbourhood in $\Delta ^o_{\beta }$ should be clear.

These  local  trivializations  fit  together  defining  a
principal bundle structure on $TG^o$, the transition maps being inherited
from  the bundle structure of $T \hookrightarrow G \rightarrow G/T$.
\hfill $\Box$

The construction above involves the choice of a Weyl chamber.
Locally  we  have
\beq
TG^o \buildrel{loc}\over\approx ( W^o\times G/T\times G/T)\times T.
\enq
By adding another copy of $W^o$ and $T$, one
is able  to  identify $TG^{o}$  with  a  quotient  of
$({\cal W}^{o}\times G)^{2}$ by some suitable relation.
We now describe this 'splitting' and 'fusion'.

Let us consider the manifold $P := W^o\times G\ni (w,g)$,
and the 1-form
\beq
\theta  := K(w, g^{-1}dg) .
\label{theta}
\enq
 Its differential $d\theta  \equiv  \omega $ defines a symplectic
 stucture, as one can check that it is non-degenerate on $P$.
To write it out more explicitly, let us locally represent
an element of $G$ as a product of the section  $\sigma$ of $G
\rightarrow G/T$ and an element $t \in T$.
Then
\begin{eqnarray}
\omega  &=&  K(dw,\wedge \sigma^{-1}d\sigma) + K(dw,\wedge t^{-1}dt)
\label{omega in K} \\
&& - K(w,\sigma ^{-1}d\sigma \wedge \sigma^{-1}d\sigma) .\nonumber
\end{eqnarray}
 The last term of the  sum  is  nothing  but  a   local
expression for a non--degenerate  symplectic  form  on the adjoint
orbit $G/T$ \cite{Kirillov} corresponding to the point $w$. The middle one
canonically couples the  Weyl chamber to the torus, and the nondegeneracy
can be read off from \eqn{sectsym}.

We now take two copies   of $(P,\omega )$, a 'left' and a 'right' copy
(distinguished by the indices  $r$  and $l$), and  introduce the
symplectic product
\beq
({\cal P} ,\omega _{\cal P}) := (P_{l}\times P_r, \omega _{l}- \omega _r).
\label{symplectic product}
\enq
The left and right sectors can be fused into the particle phase space:
\begin{quote}
{\it $TG^o $ is the symplectic reduction of $ ({\cal P},\omega _{{\cal P}})$
\\by  the constraints \mbox{$w_l = w_r$}.   } \label{reduction theorem}
\end{quote}
The constraints are solved by embedding of
$N:= W^o\times G\times G$ in ${\cal P}$ :
\beq
N\ni (w,g_{l},g_{r}) \rightarrow  i(w,g_{l},g_{r}):=
                (w ,g_{l}, w , g_{r})\in {\cal P} .
\label{33}
\enq
 The pull-back of the potential of the  symplectic form is given by:
\beq
i^{*}(\theta_{\cal P}=\theta_l-\theta_r) =
K(w, g^{-1}_{l}dg_{l}) - K(w, g^{-1}_{r}dg_{r}) .
\enq
 It is easy to check that the projection :
\beq
N \ni  (w,g_{l},g_{r}) \mapsto  \pi (w,g_{l},g_{r}) :=
(g_{l}g^{-1}_{r}, Ad_{g_r}w )\in  (G\times {\cal G})^o
\label{cfus}\enq
 satisfies
\beq
\pi ^*(\alpha ) = i^{*}(\theta _{\cal P}),
\label{37}\enq
where $\alpha $ is the  Liouville  form  on $TG^o$.\hfill $\Box$

Splitting the phase space into two sectors we are free to split the
Hamiltonian (\ref{ham}) to generate dynamics in each of them. This
procedure is by no means unique.
One can  take the 'left' ('right') Hamiltonian  simply as
$ h_{l,r} := {1 \over 4} K(w_{l,r},w_{l,r})$
 i.e. they are both proportional to the quadratic Casimir.

\end{document}